\documentclass[superscriptaddress,onecolumnpage,notitlepage,nofootinbib]{revtex4-1}

\pdfoutput=1
\usepackage{graphicx} 
\usepackage{tikz}
\usepackage{amsmath}
\usepackage{amssymb}
\usepackage{natbib}
\usepackage{color}
\usepackage{float}
\usepackage{epstopdf}
\usepackage[normalem]{ulem}
\usepackage{pgfplots,pgfplotstable}
\usepackage{soul}
\usepackage{accents}
\usepackage{nameref}
\usepackage{balance}
\usepackage[all]{nowidow}
\usepackage{wrapfig}
\usepackage{hyperref}

\usepackage{color}
\usepackage{gensymb}
\usepackage{amsmath}
\usepackage{dsfont}
\usepackage{blindtext}
\usepackage{xcolor}
\usepackage{url}
\usepackage{siunitx}
\usepackage{mathtools}
\DeclarePairedDelimiter\abs{\lvert}{\rvert}

\begin{document}

\title{Efficient snap-through of spherical caps by applying a localized curvature stimulus}

\author{Lucia Stein-Montalvo}
\email{lsmontal@bu.edu}
\affiliation{
	Department of Mechanical Engineering, Boston University, Boston, MA, 02215.
}%

\author{Jeong-Ho Lee}
\affiliation{
	Department of Mechanical Engineering, Boston University, Boston, MA, 02215.
}%

\author{Yi Yang}
\affiliation{
	Department of Mechanical Engineering, Boston University, Boston, MA, 02215.
}%

\author{Melanie Landesberg}
\affiliation{Yale University, New Haven, CT 06520.
}%

\author{Harold S. Park}
\affiliation{
	Department of Mechanical Engineering, Boston University, Boston, MA, 02215.
}%

\author{Douglas P. Holmes}
\email{dpholmes@bu.edu}
\affiliation{
	Department of Mechanical Engineering, Boston University, Boston, MA, 02215.
}%

\begin{abstract}
In bistable actuators and other engineered devices, a homogeneous stimulus (\textit{e.g.} mechanical, chemical, thermal, or magnetic) is often applied to an entire shell to initiate a snap-through instability. In this work, we demonstrate that restricting the active area to the shell boundary allows for a large reduction in its size, thereby decreasing the energy input required to actuate the shell. To do so, we combine theory with 1D finite element simulations of spherical caps with a non-homogeneous distribution of stimulus--responsive material.  We rely on the effective \textit{curvature stimulus}, \textit{i.e.} the natural curvature induced by the non-mechanical stimulus, which ensures that our results are entirely stimulus-agnostic.  To validate our numerics and demonstrate this generality, we also perform two sets of experiments, wherein we use \textit{residual swelling} of bilayer silicone elastomers--a process that mimics differential growth--as well as a magneto-elastomer to induce curvatures that cause snap-through. Our results elucidate the underlying mechanics, offering an intuitive route to optimal design for efficient snap-through.
\end{abstract}
\maketitle

\section{Introduction} \label{sect:intro}
The spherical green alga \textit{Volvox globator} swims thanks to thousands of synchronously flapping flagella. This motility is surprisingly hard-earned: after cell division, the flagella direct inward, toward the center of the sphere. Thus, the Volvox embryo must entirely invert itself during morphogenesis. 
To do so, cells in a ``bend region" adopt a wedge shape, which creates a localized curvature that leads to an eventual instability reminiscient of snap-through~\cite{Haas2015,Haas2018}. Depending on the species, the bend region either propagates from the open phialopore (located at the pole), where four lips peel back to drive type-A inversion, or begins at the equator, where invagination leads to type-B inversion~\cite{Hohn2015}. Arrested inversion of mutant Volvox has been linked to insufficient size of the bend region or intrinsic curvature therein~\cite{Haas2018}. 

Like the Volvox embryo--and indeed many other living organisms including, famously, the Venus fly trap \cite{Forterre2005}--engineers use snap-through instability of shell structures for functionality. Snapping releases stored elastic energy and does not require a continuously applied stimulus to maintain an inverted shape in bistable structures. Thus, snap-through instability is a particularly attractive mehanism for \textit{e.g.} robotic actuators or mechanical muscles~\cite{Shao2018,Gorissen2020}, optical devices~\cite{Holmes2007}, and dynamic building façades~\cite{Song2018}. Each relies on a combination of geometry-endowed bistability~\cite{Taffetani2018} and a snap-inducing stimulus to acheive its purpose. The stimulus can be mechanical--\textit{e.g.} an indentation force~\cite{Wan2021}, pressure~\cite{Gorissen2020}, or torque from a child's hands inverting a jumping popper toy~\cite{Pandey2014}--or non-mechanical, \textit{e.g.} temperature~\cite{Jakomin2010}, voltage~\cite{Shao2018}, a magnetic field~\cite{Seffen2016}, or differential growth~\cite{Goriely2017,Lee2021} or swelling~\cite{Pezzulla2018,Abdullah2016}. An important connection between these wide-ranging stimuli is that each destabilizes a shell by generating a change in the shell's curvature relative to its original, stress--free curvature.

In particular, non-mechanical stimuli alter the stress-free reference state, producing spontaneous or \textit{natural curvature}. A classic example is the bimetallic beam of Timoshenko, which curves as its two layers experience different expansive responses to an increasing thermal stimulus~\cite{Timoshenko1925}. 
The natural curvature may be observed as the shape that a beam adopts when free of external constraints and exposed to a stimulus that causes this beam to bend. In plates or shells, however, the natural curvature is generally not achievable at all points in the slender structure. In such \textit{non-Euclidean}~\cite{Efrati2009} plates and shells, geometric incompatibility leads to residual stresses and often complex reconfiguration~\cite{Klein2007,Sharon2010}. 
Still, the relationship between a stimulus and the corresponding natural curvature it induces in a residually-stressed structure can be discerned \textit{via} simple experimental methods based on the bending deformation of an equivalent beam. Calibration can be performed as an independent experiment, or in a manner similar to the \textit{opening angle method}, wherein an axial slice exposes stresses induced by differential growth in tubular biological structures like arteries~\cite{Alastru2007}. Thus, natural curvature serves as useful ``effective" stimulus, allowing for generalization of these many non-mechanical, curvature-inducing stimuli. This concept was recently formalized within a non-Euclidean theoretical framework based on Koiter shell theory, revealing that the \textit{curvature stimulus} behaves like a mechanical potential~\cite{Pezzulla2018,Holmes2020}. 

Advanced functional devices actuated by curvature-inducing stimuli often require significant energy input, or have high material costs. Common commerically available dielectric elastomer (DE) films require up to  \SI{150}{\volt \per \micro \meter} for most applications, resulting in driving voltages in the kV range, and proportionally high energy consumption~\cite{OHalloran2008,Lu2020}. Using less of these smart materials can decrease both cost and the likelihood of device failure. This need has inspired custom production of ultrathin electroactive films via methods like pad-printing~\cite{Poulin2015} or electrospraying~\cite{Weiss2016}. Clearly, it is desirable to reduce these costs for DEs and other stimulus-responsive devices without requiring such efforts. As the localized bend region of the developing Volvox suggests, snap-through behavior may be preserved in spherical caps while the size of the active region--that is, the portion of the shell subjected to an arbitrary curvature-inducing stimulus--is significantly reduced. 

In the present work, we show that when the active portion is strategically placed, the magnitude of the stimulus need not increase, thus allowing for significantly reduced energy and material needs overall. We demonstrate this with 1D numerical simulations performed in COMSOL Multiphysics, wherein a curvature stimulus acts on a section of an otherwise passive shell. We validate our numerics with experiments, in which shells respond to different curvature-inducing stimuli--this also serves to demonstrate the generality of our findings. In one series of experiments, we rely on localized differential swelling of silicone elastomers, and in the second we use a magneto-active elastomer. Our numerical and experimental methods are described in Sect.~\ref{sect:methods}. In Sect.~\ref{sect:polevsedge} we present our results to answer the question: Is the active region more effective when placed in the bulk of the shell, or at the edge? Next, we ask: What is the the ideal size of the active region? In other words, what is the smallest active region that preserves snap-through behavior, without requiring the magnitude of the stimulus to increase? In Sect.~\ref{sect:optimize} we present a scaling solution for the ideal size of the active region, derived \textit{via} energy minimization. These findings, which our data support, offer a mechanics-informed route to optimization. The underlying physics revealed prompt an additional examination of how the critical curvature scales with the active area in each configuration (Sect.~\ref{sect:scalings}), which adds rigor to the comparison presented earlier (in Sect.~\ref{sect:polevsedge}). We offer concluding remarks in Sect.~\ref{sect:conclusion}.

\section{Methods} \label{sect:methods}
\begin{figure}[h!]
	\centering
	\includegraphics[width=\linewidth]{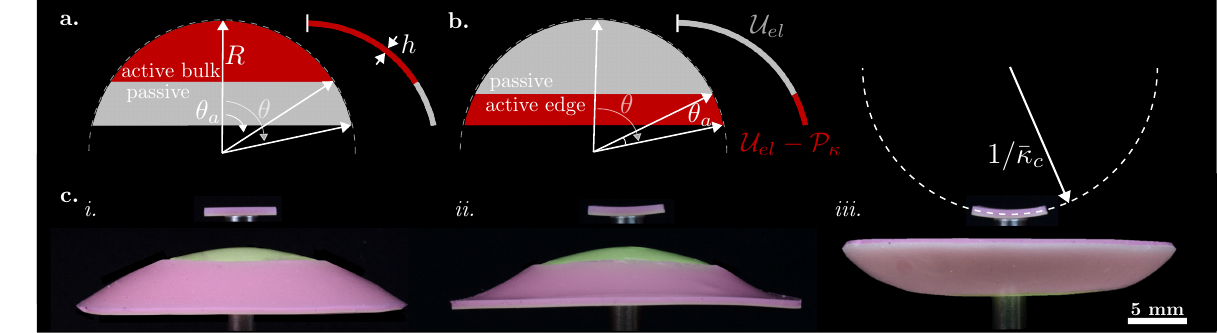}
	\caption{Schematics of the a. active bulk and b. active edge configurations of the partially active spherical cap, with corresponding profile curves used in FEM simulations. Relevant parameters are labeled: the angular shell depth $\theta$ measured from the pole, the angular size of the active (red) region $\theta_a$, and the radius of curvature $R$ and thickness $h$. As depicted in b., the COMSOL model minimizes the elastic energy $U_{el}$ over the whole body, with an additional contribution from the curvature potential $\mathcal{P}_{\kappa}$ in the active region. c. Images from a typical residual swelling experiment in the active edge configuration. In \textit{i.}, the PVS beam (above), cut from the same cast as the partial bilayer shell (below) have just cured. The beam is free to adopt the curvature $\bar{\kappa} = 1/R + \kappa$, revealing the evolving natural curvature $\kappa$ in active region of the shell. \textit{ii.} In the bilayer region at the edge, the outer pink layer contracts while the green layer beneath it expands due to residual swelling. The resulting curvature (seen in the beam) creates a torque-like effect, causing the boundary of the shell to curl upward. The curvature evolves to a critical value $\bar{\kappa}_c$ when snap-through occurs, as shown in \textit{iii.} 
	}
	\label{fig:schematics_sampleresults}
\end{figure}

In this work, we study spherical caps in two configurations: ``active bulk" (Fig.~\ref{fig:schematics_sampleresults}a), and ``active boundary" (Fig.~\ref{fig:schematics_sampleresults}b). We refer to the angular depth of the shell measured from the pole as $\theta$. The angular extent of the active region is denoted by $\theta_a$, and is measured either from the pole or from the end of the passive region for the active pole and edge configurations, respectively. The shell thickness is given by $h$ and the radius of curvature by $R$. Below, we describe our numerical and experimental methods. 

\subsection{1D numerics} \label{sect:numerics}
We performed finite element simulations in COMSOL Multiphysics 5.2. Following Refs.~\cite{Pezzulla2018,Holmes2020}, and because our experiments (Sect.~\ref{sect:exp}) confirm that the shells we study retain rotational symmetry at least up to the point of snapping, the energy is minimized in the 1D profile curve of the shell. A circular segment with radius of curvature $R \in [9.65,45.25]$ mm represents the midline of a spherical cap with an angular depth $\theta \in [0.52, 1.14]$ rad measured from the pole. This corresponds to $\theta/\sqrt{h/R} \in [2.2,9.6]$, which covers most the range where curvature-induced snap-through occurs for fully active shells~\cite{Pezzulla2018}. As we discuss in Sect.~\ref{sect:optimize}, this parameter compares the depth of the shell to the characteristic size of its bending-dominated boundary layer, which becomes central to our analysis. The ratio $\theta_a/\theta$ is between $0.3$ and $1$ for the data presented herein, because we did not observe snapping for shells with $\theta_a/\theta<0.3$. The shell thickness $h \in [0.5,1.3]$ mm (so that $h/R \in [0.015,0.135]$) enters in the energy. The material is linear, elastic, isotropic, and incompressible with Poisson's ratio $\nu \in [0.25,0.5]$. The Young's Moduli $E_a$ and $E_p$ of the active and passive regions were set to either $E_a = E_p \in [0.1,20]$ MPa, or $E_a/E_p \in [0.005,200]$. Despite that we assume a homogenous modulus (see Sect.~\ref{sect:optimize}), we observe no significant impact of the modulus on our results. A Dirichlet boundary condition is applied to the endpoint at the pole, while the edge end remains free. 

Our model relies on the theoretical framework recently introduced by Pezzulla, {\em et al.}~\cite{Pezzulla2018} and detailed further by Holmes {\em et al.}~\cite{Holmes2020}, which offers significant advantages for numerical energy minimization in the presence of non-mechanical stimuli. The authors demonstrate that in the absence of in-plane stretching of the middle surface, the curvature-inducing stimulus may be decoupled from Koiter's elastic energy and applied as a potential of the natural curvature $\kappa$. Thus, it is straightforward to selectively apply a curvature stimulus. The elastic energy ($\mathcal{U}_{el}$ in the schematic in Fig.~\ref{fig:schematics_sampleresults}b) must be minimized over the entire body, while the curvature potential $\mathcal{P}_{\kappa} \sim \kappa$ contributes to the total energy only in the active region. To capture the sudden, nonlinear snap-through instability, we use a custom arc-length method to vary the curvature stimulus $\kappa$ ~\cite{Pezzulla2018}. A complete derivation of the equations used in COMSOL is provided in Ref.~\cite{Holmes2020}. 

\subsection{Experiments} \label{sect:exp}
To validate our numerical results, we performed experiments in which shells are made, in part, of stimulus-responsive elastomers. To emphasize the generality of the curvature stimulus, we did this in two ways. In one set of experiments we used \textit{residual swelling} of bilayer silicone elastomers~\cite{Pezzulla2015,Pezzulla2016,Pezzulla2018,Stein-Montalvo2019}, wherein diffusion of free polymer chains causes a geometric incompatibility resolved by curving. Deformation from residual swelling is growth-like and permanent. In additional experiments, we used magneto-active elastomers, which reversibly curve in response to a magnetic field. Below, we describe the methods used for each type of experiment.

\subsection{Residual swelling experiments}\label{sect:residswell}
It was previously demonstrated that residual swelling creates a curvature sufficient to drive inversion of full bilayer spherical caps~\cite{Pezzulla2018}. In the present work, only a portion of the shell is subjected to residual swelling. Within this active region, we layer two polyvinylsiloxane (PVS) elastomers ($\nu = 0.5$), which we refer to as \textit{green} (Zhermack Elite Double 32, E=0.96 MPa) and \textit{pink} (Zhermack Elite Double 8, E=0.23 MPa). In a procedure detailed in Appendix~\ref{sect:shellfab}, we cast the materials one-by-one in their fluid state over a metal ball bearing to form a nominally green cap with either a bilayer ring at the edge, or a bilayer cap at the pole. In the active region, the outer layer is pink (see Fig.~\ref{fig:schematics_sampleresults}c). After crosslinking (which occurs in about 20 minutes at room temperature), the pink elastomer is left with residual, uncrosslinked polymer chains. Thus, in the bilayer region where the pink and green material are in contact, there is a concentration gradient of free polymer chains. To resolve this, chains flow from the pink to the green elastomer. As such, the pink region loses mass, or shrinks, while the green region grows. When this differential swelling occurs in an initially flat beam, the structure curves to accommodate the geometric mismatch, adopting the (evolving) natural curvature. In our spherical caps, the bilayer region around the boundary curls upward, forming a lip like that which appears when a jumping popper toy is inverted~\cite{Pandey2014,Taffetani2018}. Geometric constraints prevent the swelling bilayer cap from achieving its natural curvature, and the structure develops residual stresses as a result. 

It is necessary that we measure the natural curvature in order to discern the magnitude of the curvature stimulus at a given time, which is central to our analysis. To do so, for the active edge experiments, we slice a small, vertical (initially curved) beam from the material on the ball-bearing just below where the cut is made at the base of the spherical cap. The free, unconfined beam adopts a curvature $\bar{\kappa}$, which is the sum of the initial curvature $-1/R$ and the natural curvature $\kappa$. Thus, the (natural) curvature stimulus is $\kappa = \bar{\kappa} + 1/R$~\cite{Pezzulla2015,Pezzulla2016,Pezzulla2018}. We mount the shell and beam side-by-side and use a Nikon D610 DSLR Camera to take time-lapse images at a rate of one photo per minute. As residual swelling is a diffusive process, the time to deform scales with the square of the dimension across which swelling occurs~\cite{Pezzulla2015}. In our experiments, that is the thickness $h$, and these shells reach maximum deformation by about two hours post-cure. The critical curvature $\kappa_c$ is identified in ImageJ by fitting a circle to the beam in the image when snap-through occurs (see Fig.~\ref{fig:schematics_sampleresults}c,\textit{iii}.)--the radius of curvature is $1/\bar{\kappa}_c$.  For the active bulk configuration, the beam is cut from the center of the shell immediately after snapping. Thus, unlike in the active boundary configuration wherein we track the natural curvature throughout deformation, only the critical curvature $\kappa_c$ is obtained for these experiments. Residual swelling experiments produce curvatures up to $\kappa<\frac{1}{4h}$~\cite{Pezzulla2015,Pezzulla2016}. Deep shells require higher curvatures for snap-through~\cite{Pezzulla2018}, so the geometric range accessed in experiments is limited compared to simulations. We performed experiments with $h/R \in [0.02, 0.07]$, $\theta \in [0.58, 0.98]$ rad, and $\theta_a/\theta \in [0.38, 1]$. 

\subsection{Magneto-elastomer experiments}\label{sect:magneto}
The magneto-active shells are fabricated using a similar bilayer casting approach to the residual swelling shells. In this case, the ferromagnetic active layer consists of iron oxide ($Fe_3O_4$) nanoparticles (Sigma-Aldrich 637106) mixed with green PVS at a weight ratio of 20$\%$. A passive PVS cap is made first, followed by a ferromagnetic edge ring. Then, an additional PVS layer is added to join the two, resulting in a shell with relatively uniform thickness with a passive bulk and active edge. We did not study the active bulk configuration with magneto-elastomer shells. A $25.4$ mm cubic NdFeB magnet (N52 grade by SuperMagnetMan) is used to generate the magnetic gradient. The strength is measured by the magnetic flux density $\mathcal{B}$, using a magnetometer (PCE Instruments Inc). The relative permeability and remanence of the NdFeB magnet are $1.04$ and $1.45$, respectively, giving $\mathcal{B}=0.562$ T at the surface center of the magnets.

We tested five magnetic shells with with $h/R \in [0.02,0.04]$, $\theta = 0.86$ rad, and an active boundary region such that $\theta_a/\theta \in [0.07,1]$. The center of the magneto-active shell is mounted on a PVS column (diameter $4.5$ mm) to support the shell when magnetic field is applied. The NdFeB magnet is mounted vertically on Instron 5943 with a $5$N load cell, then quasi-statically approaches the shell at a rate of $0.1$ mm/s using the software Bluehill3. When the magnet reaches a critical distance from the shell, the magnetic body force acting on the ferromagnetic edge triggers snapping. The critical displacement is retrieved from the force-displacement curve. In order to find the corresponding critical $\mathcal{B}$, we simulate the magnetic field using the AC/DC module in COMSOL 5.6. This allows us to visualize the magnetic field and obtain $\mathcal{B}$ at given spatial points. Although the magnetic stimulus is nonhomogenous, we select the value of $\mathcal{B}$ at the edge of the shell for our analysis. To relate $\mathcal{B}$ to the curvature stimulus $\kappa$, we perform an additional set of experiments in which we measure the curvature in a ferromagnetic beam, oriented in the same direction as the edge of the shell, as the magnet approaches. For details on the magnetic field visualization and the magnetic flux-curvature calibration, see Appendix~\ref{sect:flux-curvature}.

\section{Comparison of active bulk and boundary}\label{sect:polevsedge}
\begin{figure}[h!]
	\centering
	\includegraphics[width = 0.5\linewidth]{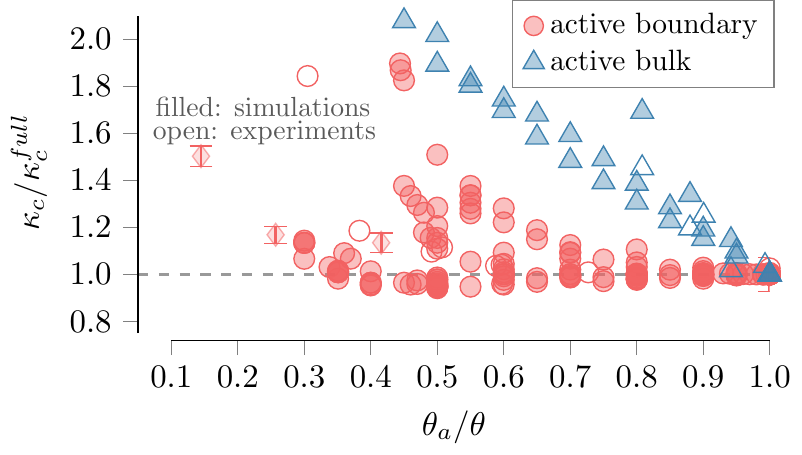}
	\caption{Comparison of the critical curvature $\kappa_c$ compared to that for a fully active shell, $\kappa_c^{full} = \kappa_c(\theta_a = \theta)$, for varied sizes of the active region in the active bulk (blue triangles) and active boundary (pink circles) configurations. Our results show a clear preference for the active boundary configuration: $\theta_a/\theta$ can be reduced to as low as about $0.35$ without requiring a higher curvature stimulus for snap-through, whereas if the bulk is active (and the boundary is passive), $\kappa_c/\kappa_c^{full} > 1$ for any $\theta_a/\theta < 1$. Error bars correspond to one standard devation, and are smaller than markers in all cases except the magneto-active experiments.}
	\label{fig:bulkvsboundary}
\end{figure}

In Fig.~\ref{fig:bulkvsboundary}, we compare the critical curvature $\kappa_c$ to that for a fully active shell, $\kappa_c^{full} \equiv \kappa_c(\theta_a=\theta)$, for varied proportions of the angular active area to the total angular depth of the shell, $\theta_a/\theta$. In the active bulk configuration, a reduction of only about $5$ percent in $\theta_a/\theta$ causes the critical curvature $\kappa_c$ to increase significantly above $\kappa_c^{full}$. Meanwhile, we observe that when the boundary is active and the passive region lies in the bulk, the active portion can be reduced as much as $65$ percent in some cases without requiring $\kappa_c/\kappa_c^{full} > 1$ for snap through. 

As discussed in Sect.~\ref{sect:numerics}, it has been shown that a stimulus which induces intrinsic curvature may be treated as a \textit{curvature potential}, which is decoupled from Koiter's elastic energy~\cite{Pezzulla2018}. Pezzulla \textit{et al.} demonstrate that this curvature potential may be further decomposed into two work-like terms, \textit{i.e.} $\mathcal{P}_{\kappa} = -\mathcal{W}_{bulk}- \mathcal{W}_{edge}$, which behave like an applied pressure in the bulk and a torque on the boundary. The pressure-like term scales as:
\begin{equation}\label{pressure}
\mathcal{W}_{bulk} \sim h^4 \kappa^2 (1-\cos \theta),
\end{equation}
and the torque-like term as:
\begin{equation}\label{torque}
\mathcal{W}_{edge} \sim h^4 \kappa^2 \bigg(\frac{R}{h}\bigg)^{3/2} \sin \theta.
\end{equation}
Comparing the two contributions gives 
\begin{equation}\label{compare}
\abs*{\frac{\mathcal{W}_{edge}}{\mathcal{W}_{bulk}}}\sim \frac{(R/h)^{3/2}}{\tan(\theta/2)} >> 1,
\end{equation}
showing that the edge work dominates in thin shells~\cite{Pezzulla2018}.

The implication explains our findings in general: removing a portion of active area from the edge weakens the effect of the curvature stimulus more than if the same amount of area were removed from the bulk. Conversely, the inequity suggests that the active edge configuration is the more efficient choice for the design of structures that will snap with minimal energy needs. To clearly distinguish between these regimes, henceforth we will refer to the critical curvature in the active boundary configuration as $\kappa_c^{edge}$, and that in the active bulk configuration as $\kappa_c^{bulk}$. Despite this clear preference for an active boundary, the optimal size of the active region is not clear from the scattered data in Fig.~\ref{fig:bulkvsboundary}. We investigate this in the following section.

\section{Optimal size of active boundary \textit{via} energy minimization}\label{sect:optimize}
\begin{figure}[h!]
	\centering
	\includegraphics[width = 0.58\linewidth]{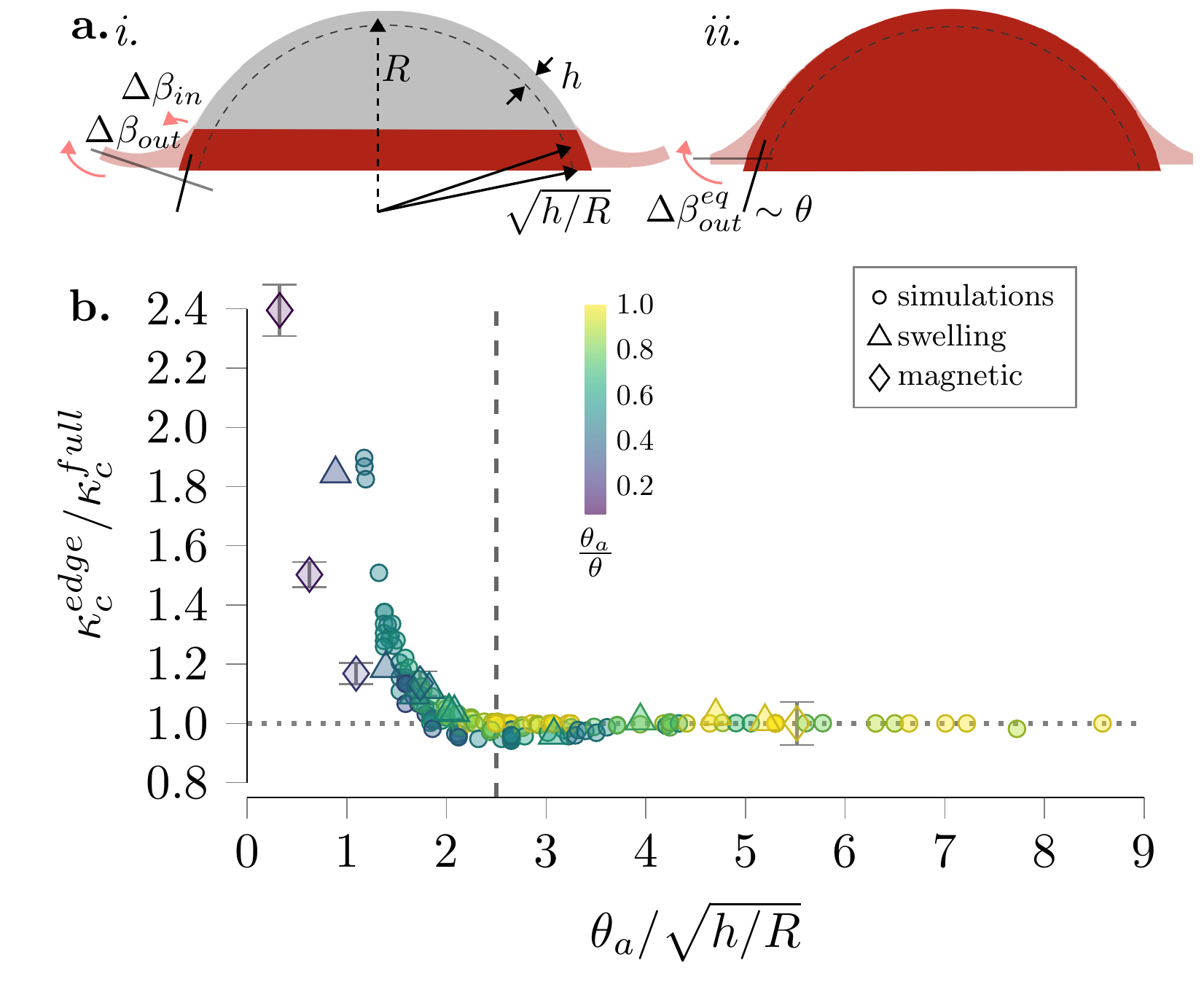}
	\caption{a. Schematic of the energy minimization scheme. The partially active shell is shown in \textit{i.}, where $\kappa$ delivers a torque-like effect (pink arrows) along the free boundary and at the active-passive interface, which drives the tangent to the edge (black line) to rotate an amount $\Delta \beta_{out}$ (to the dark gray line) and that at the interface a smaller amount $\Delta \beta_{in}$. The characteristic angular size of the natural boundary layer, $\sqrt{h/R}$ is labeled. \textit{ii.} The equivalent fully active shell, which feels a torque from the curvature stimulus only about the free edge. Fully active shells snap-through when the tangent to the edge becomes approximately horizontal~\cite{Pezzulla2018}. b. The normalized critical curvature for the active edge configuration \textit{vs.} the active area $\theta_a$ compared to the characteristic length of the natural boundary layer, $\sqrt{h/R}$. The data collapse supports the result Eq.\eqref{thetaa_scaling}, \textit{i.e.} the optimal size for the active boundary region $\theta_a^*$ scales with the geometry-endowed boundary layer size.}
	\label{fig:edge_kappac}
\end{figure}

To discern the most efficient size of the active region at the shell boundary, we minimize the total potential energy in the system $\mathcal{U}$, which consists of the internal elastic energy, and the non-mechanical loading due to curvature~\cite{Pezzulla2018,Holmes2020,Lee2021}. In the latter, the torque-like contribution at the boundary dominates over the pressure-like effect in the bulk. This effect is amplified in the active edge configuration, as some or all of the bulk is passive. Accordingly, we observe very small bulk deformation compared to boundary rotation (see Fig.~\ref{fig:angles}b.) As such, we neglect the pressure-like contribution in the bulk, but include the additional torque that arises at the boundary between the active and passive regions (Fig.~\ref{fig:edge_kappac}a). This gives:
\begin{equation}
\label{shellenergy}
\mathcal{U} = \mathcal{U}_{K} - \oint_{out}{B(1+\nu)\kappa(\Delta\beta_{out})} \: d\mathring{s} - \oint_{in}{B(1+\nu)\kappa(-\Delta\beta_{in})} \: d\mathring{s}
\end{equation}
where $\mathcal{U}_{K}$ is the Koiter shell energy~\cite{Koiter1973}, $B$ is the bending rigidity, $\Delta\beta_{out}$ and $\Delta\beta_{in}$ are the angle change at the boundary and the active-passive interface, respectively, and $d\mathring{s}$ is the line element.  The second and third terms quantify the torque-like contribution of the non-mechanical loading induced by the natural curvature at either end of the active edge region. Note that a positive angle change corresponds to a counter-clockwise rotation.

Because we expect axisymmetric deformation up to snap-through, these non-mechanical loading terms can be collected into the free boundary line integral as
\begin{equation}
\label{shellenergyeq}
\mathcal{U} = \mathcal{U}_{K} - \oint_{out}{B(1+\nu)\kappa\Delta\beta_{out}^{eq}} \: d\mathring{s}
\end{equation}
where $\Delta\beta_{out}^{eq} = \Delta\beta_{out} - \Delta\beta_{in}\left(1-\frac{\theta_{a}}{\theta}\right)$. Physically, this corresponds to transforming the shell into an equivalent one that experiences only the angle change $\Delta\beta_{out}^{eq}$ at the boundary, and is fully active, \textit{i.e.} the natural curvature $\kappa$ acts on the whole body (see Fig.~\ref{fig:edge_kappac}b). Then, from Ref.~\cite{Pezzulla2018}, we can postulate that the snapping occurs when the colatitude-direction tangent vector at the boundary of the equivalent shell approximately becomes horizontal, \textit{i.e.} at $\Delta\beta_{out}^{eq} \sim \theta$. Note that this does not imply that $\Delta\beta_{out} \sim \theta$, which as we discuss in Sect.~\ref{sect:scalings}, is not necessarily the case. 

For thin surfaces where the non-mechanical stimulus acts through-the-thickness, the contribution of the mid-surface stretch induced by the stimulus can be neglected~\cite{Pezzulla2018,Holmes2020}. Then, assuming the colatitude-direction bending strain is much larger than the azimuthal one, the shell energy  $\mathcal{U}$ scales as
\begin{equation}
\label{shellenergyscaling}
\mathcal{U} \sim \frac{B}{2}\left(b_{1}^{1}\right)^{2}_{p}A_{p} + \frac{B}{2}\left(b_{1}^{1}\right)^{2}_{a}A_{a} - B(1+\nu)\kappa\left[\Delta\beta_{out} - \Delta\beta_{in}\left(1-\frac{\theta_{a}}{\theta}\right)\right](2\pi R \theta)
\end{equation}
where $b_{1}^{1}$ is the characteristic curvature in the colatitude-direction and the subscripts $p$ and $a$ denote the passive bulk and active edge regions. The curvature $b_{1}^{1}$ can be estimated as the angle change along an arc over its length, so that 
\begin{subequations}
	\begin{equation}\label{b11p_scaling}
	(b_{1}^{1})_{p} \sim \frac{-(\theta - \theta_{a} - \Delta\beta_{in})}{R(\theta-\theta_{a})}
	\end{equation}
	\begin{equation}\label{b11a_scaling}
	(b_{1}^{1})_{a} \sim \frac{-(\theta_{a}+\Delta\beta_{in}-\Delta\beta_{out})}{R\theta_{a}}.
	\end{equation}
\end{subequations}
The active area $A_{a}$ scales as $2\pi R^{2} \theta \theta_{a}$, and the passive area $A_{p}$ as $\pi R^{2}\theta^{2} - A_{a}$.

Inserting these scalings and minimizing Eq. \eqref{shellenergyscaling} with respect to $\Delta\beta_{in}$ and $\Delta\beta_{out}$ gives the solutions for these angle changes in the deformed configuration. Evaluating the result at the presumed point of instability, $\Delta\beta_{out}^{eq} \sim \theta$, gives the critical natural curvature of the partially active shell at snapping as: 
\begin{equation}
\label{kappacedge}
\kappa_{c}^{edge} \sim \frac{\theta}{\theta_{a}(1+\nu)R}.
\end{equation} 
Assuming the minimum critical curvature occurs at $\kappa_c = \kappa_{c}^{full}$, where from Ref.~\cite{Pezzulla2018}:
\begin{equation}\label{MPscaling_kappac}
\kappa_{c}^{full} \sim \frac{\theta}{(1+\nu)\sqrt{Rh}},
\end{equation}
we find that the optimal size of $\theta_{a}$, which we denote as $\theta_a^* $, should scale as:
\begin{equation}\label{thetaa_scaling}
\theta_a^* \sim \sqrt{h/R}.
\end{equation}
This result is shown in Fig.~\ref{fig:edge_kappac}b. We see a promising collapse of our data, and a minimum emerges at about $2.5$ on the horizontal axis, indicating that the true optimal $\theta_a^* \approx 2.5 \sqrt{h/R}$ for our shells, which vary in geometry, $E$, and $\nu$. We note that for moderately deep shells, we observe critical curvatures even below the predicted $\kappa_c^{full}$ (as low as $\kappa_c = 0.95 \kappa_c^{full}$, for a shell depth $\theta/\sqrt{h/R} = 5.3$ when $\theta_a/\theta = 0.5$). We suspect that this results from additional destabilizing effects delivered by the torque at the active-passive interface, which tends to help flatten the bulk of the shell. We note that magneto-elastomer shells carry more error than the residual swelling experiments and simulations. This is largely due to the non-uniform magnetic gradient generated from the NdFeB magnet. Additionally, the shells are highly sensitive to fabrication errors, which can result in a nonuniform ferromagnetic boundary, when $\theta_a$ is small.

The result of our energy minimization scheme given in Eq.~\eqref{thetaa_scaling} coincides with the angular length scale of the bending-dominated \textit{boundary layer}~\cite{Fung1955,Niordson1985} for a spherical cap, $\sqrt{h/R}$. In open shells, the characteristic length of the boundary layer is that which is close enough to the edge that the thickness ``feels" relatively large in comparison~\cite{Efrati2009b}. In this region the energy balance shifts (thin shells are nominally stretching energy-dominated, preferring isometry), and so does the preferred deformation mode. The boundary layer is readily observed as the lip that curls upward when a spherical cap of finite thickness, \textit{e.g.} a tennis ball sliced in half, is everted~\cite{Taffetani2018,Holmes2019}. In this region, the shell ``tries" to return to its initial state, resolving some of the stress that arises due to stretching above the midline and compression below it. Only if the boundary layer is large compared to the shell depth is the everted shell unstable, snapping back to its undeformed configuration. If the boundary layer is relatively small, the shell is bistable and can rest in its everted state indefinitely. 

This scenario appears to be analogous to the non-everted, partially active shells we study: if the bending region is large, snap-through occurs easily, whereas a shell with a small active region is stable even against a higher stimulus. This points to an intuitive interpretation of our results: generally speaking, in the active edge configuration, if the active area meets or exceeds the boundary layer, the region predisposed to bending may do so. As a result, snapping is unaffected compared to a fully active cap. For $\theta_a < \theta_{bl}$, we interfere with the boundary layer--this can be seen as reducing the size of the effective boundary layer--and higher curvatures are required to drive snap-through. The efficacy of the geometric design, in sum, depends on how the effective boundary layer set by $\theta_a$, where we \textit{impose} bending, relates to the natural boundary layer that scales as $\sqrt{h/R}$, where the shell \textit{prefers} bending. In the active bulk configuration, any $\theta_a < \theta$ disturbs the boundary layer, forcing the critical curvature upward. Thus, this interpretation also clarifies what we saw in Sect.~\ref{sect:polevsedge}. This prompts us to briefly revisit our comparison of the active bulk and boundary configurations.

\section{Scalings for the critical curvature based on fully active shells}\label{sect:scalings}
In light of the importance of the boundary layer discussed in Sect.~\ref{sect:optimize}, we may add rigor to our claim that the active boundary configuration is more efficient than the active bulk (Sect.~\ref{sect:polevsedge}). To do so, we rely  on the established result~\cite{Pezzulla2018} (from which we determined Eq.~\eqref{MPscaling_kappac}) that the critical curvature for fully active shells scales as: 
\begin{equation}\label{MPscaling}
\kappa_c^{full} R \sim \frac{\theta}{(1 + \nu) \sqrt{h/R}}.
\end{equation}
This finding leans on the assumption that the pressure-like effect of curvature in the bulk may be negelected due to the dominant effect of curvature on the boundary, which we have similarly employed in Sect.~\ref{sect:optimize}. Additionally, it assumes based on empirical observations that the tangent vector to the boundary becomes approximately horizontal at the point of snap-through~\cite{Pezzulla2018,Lee2021}. 
\begin{figure}[h!]
	\centering
	\includegraphics[width = 0.6\linewidth]{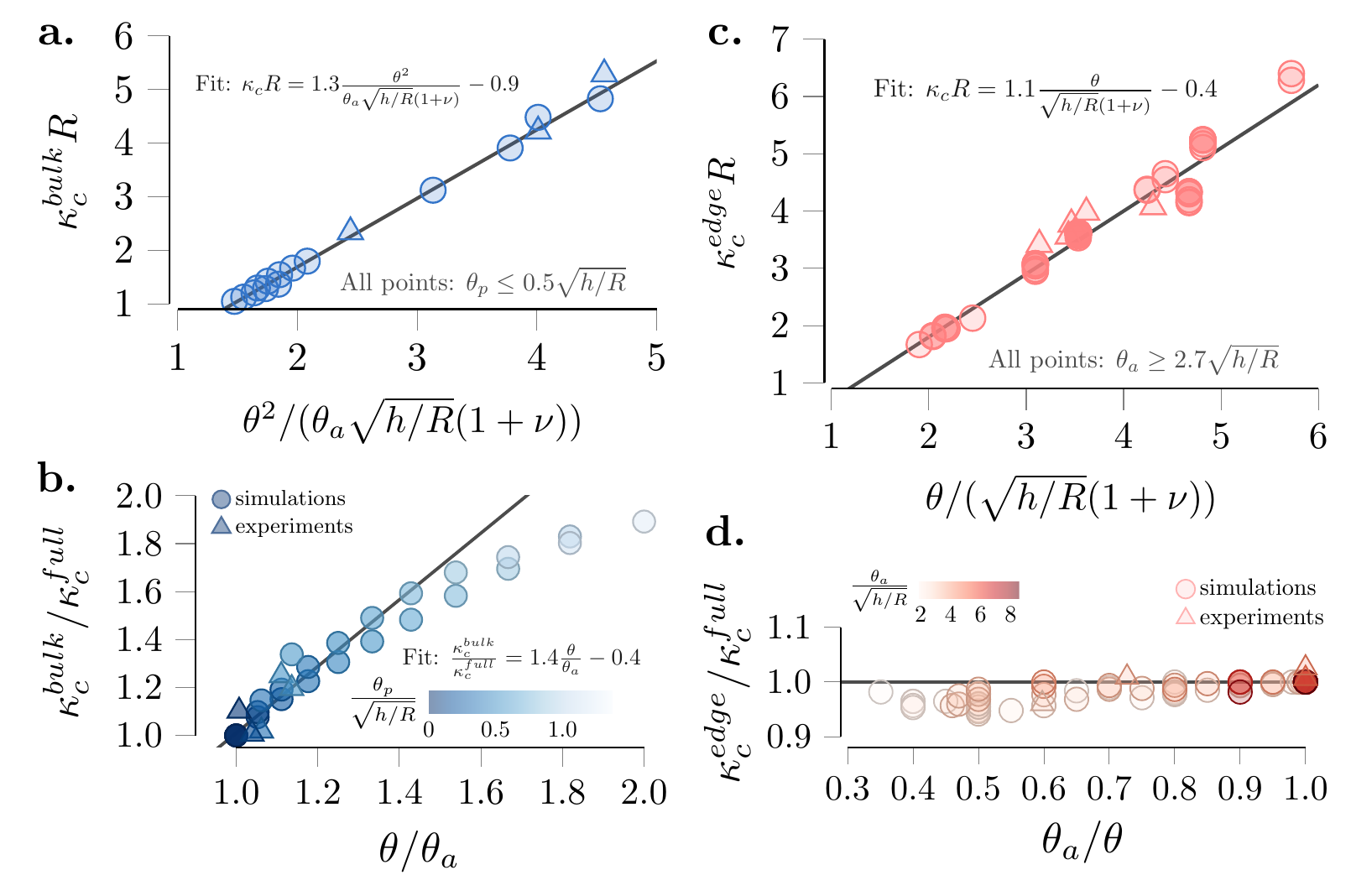}
	\caption{a. When the active portion of the shell is in the bulk, the size of the effective boundary layer is reduced. As a result, the critical curvature increases according to Eq.~\eqref{bulk_fullscaling}, and any reduction in $\theta_a$ (greater than $1$ on the x-axis) increases the the curvature stimulus $\kappa_c^{bulk}$ required for snap-through, as shown in b. As Eq.~\eqref{bulk_scaling} predicts, the stimulus increase scales linearly with $\theta/\theta_a$ when the passive edge region is smaller than the boundary layer size (darker blue points, $\theta_p/\sqrt{h/R} \lessapprox 0.5$; Fit corresponds to points in this range.) c. If instead the boundary of the shell experiences the curvature stimulus, as long as $\theta_a$ exceeds the boundary layer, the behavior follows that of the fully active shell, \textit{i.e.} Eq.~\eqref{edge_scaling}~\cite{Pezzulla2018} holds. d. As such, $\theta_a$ can be reduced to as much as $0.35 \theta$ before the critical curvature increases above that for a fully active shell, $\kappa_c^{full}$, in agreement with Eq. \eqref{edge_fullscaling}.}
	\label{fig:scalings}
\end{figure}

In the active bulk configuration, we reduce size of the effective boundary layer. Accordingly, we observe that the tangent to the edge of the effective boundary layer--that is, the active-passive interface, and not the edge of the shell--becomes approximately horizontal at the point of snap-through. Assuming the effective angular width of the boundary layer scales as $(\theta_a/\theta)\sqrt{h/R}$, it follows from Eq.~\eqref{MPscaling} that:
\begin{subequations}
	\begin{equation}\label{bulk_fullscaling}
	\kappa_c^{bulk} R \sim \frac{\theta^2}{\theta_a (1+\nu) \sqrt{h/R}},     \quad  \quad \theta_p < \theta_{bl}
	\end{equation}
	\begin{equation}\label{bulk_scaling}
	\frac{\kappa_c^{bulk}}{\kappa_c^{full}} \sim \frac{\theta}{\theta_a},     \quad  \quad \theta_p < \theta_{bl}
	\end{equation}
\end{subequations}
where $\theta_p = \theta-\theta_a$ represents the passive portion of the shell. Eqs.~\eqref{bulk_fullscaling} and \eqref{bulk_scaling} are shown in Fig.~\ref{fig:scalings}a\&b, respectively, where indeed we see that the critical curvature increases linearly with $\theta/\theta_a$ until the passive region reaches the approximate size of the boundary layer. At this point, where the entire boundary layer is inactive, the assumption that the edge work outweighs the bulk work breaks down and the critical curvature diverges.

\begin{figure}[h!]
	\centering
	\includegraphics[width = 0.52\linewidth]{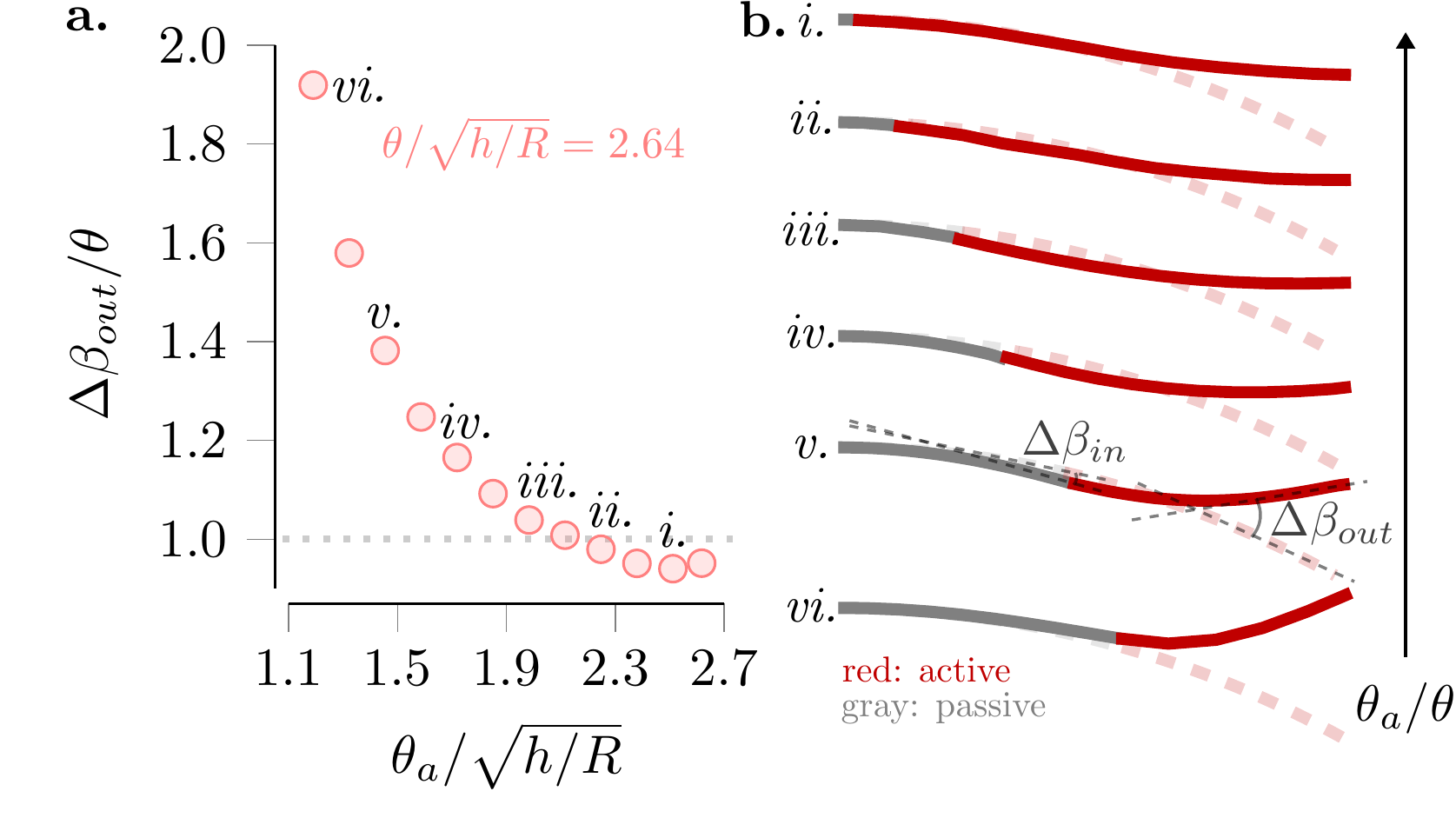}
	\caption{a. When the active area is large enough compared to the characteristic size of the boundary layer, the edge tangent requires a rotation $\Delta \beta_{out} \sim \theta$ to snap through, which results in an approximately horizontal tangent. Data is from simulations of a fixed shell geometry with changing active area. For $\theta_a \lessapprox 2 \sqrt{h/R}$ in this case, the tangent undergoes additional rotation to trigger instability. Roman numerals correspond to b., which shows the the initial (dashed) and immediately pre-snap (solid) profile curves for $\theta_a/\theta = 0.95, 0.85, ... , 0.45$. The rotations $\Delta \beta_{out}$ and $\Delta \beta_{in}$ are shown in \textit{v.} The angle change $\Delta \beta_{out}$ increases from $0.35$ rad in \textit{i.} to $0.82$ rad in \textit{vi.}}
	\label{fig:angles}
\end{figure}

For completeness, we also study the implications of Eq.~\eqref{MPscaling} on shells with active boundaries. In Sect.~\ref{sect:optimize} we used the horizontal tangent assumption ($\Delta\beta_{out}^{eq} \approx \theta$) for the equivalent fully active shell. We observe that for shells in the active edge configuration, this assumption breaks down, \textit{i.e.} $\Delta\beta_{out} > \theta$, if $\theta_a$ approaches the size of the boundary layer (see Fig.~\ref{fig:angles}). With this in mind, as long as the coincident conditions that boundary layer is in tact and the boundary tangent is approximately horizontal at snap-through ($\theta_a \gg \sqrt{h/R}$), due to the dominance of the edge work we expect no change to the critical curvature from $\kappa_c^{full}$. That is, 
\begin{subequations}
	\begin{equation}\label{edge_scaling}
	\kappa_c^{edge} R \sim \frac{\theta}{(1+\nu) \sqrt{h/R}}, \quad \quad \theta_a > \theta_{bl} 
	\end{equation}
	\begin{equation}\label{edge_fullscaling}
	\frac{\kappa_c^{edge}}{\kappa_c^{full}} =1, \quad \quad \theta_a > \theta_{bl}
	\end{equation}
\end{subequations}
Note that Eq.~\eqref{edge_fullscaling} is complementary to the result Eq.~\eqref{kappacedge}. Eqs.~\eqref{edge_scaling} and \eqref{edge_fullscaling} are shown in Fig.~\ref{fig:scalings}c\&d, respectively. Comparing Eqs.~\eqref{bulk_scaling} and \eqref{edge_fullscaling} confirms that for any small to moderate reduction of the active area, the active bulk configuration requires a higher curvature stimulus than the active boundary. 

\section{Conclusion}\label{sect:conclusion}
The aim of the present work was to guide the design of efficient snapping structures, which simultaneously minimize the active area and the magnitude of stimulus needed. We studied partially active spherical caps in two configurations--active bulk and active boundary--with a combination of theory, 1D finite element simulations, and experiments. Shells respond to a non-mechanical, curvature-inducing stimulus in the designated active region, but are passive elsewhere. 

Our mechanics-informed approach uncovered an analogy to the bending-dominated boundary layer in inverted spherical caps. This offered an intuitive interpretation of our work: Selectively applying curvature amounts to setting the size of an effective boundary layer. Like for inverted, passive spherical caps, the size of the (effective) boundary layer is closely tied to stability~\cite{Taffetani2018}. Further, the location and size of the imposed bending region determines whether it competes against or cooperates with the geometric boundary layer, wherein the shell inherently ``wants" to bend. With this view, the design principles that follow are straightforward. In the active cap configuration, some or all of the boundary layer is made passive, making snap-through harder to acheive. As a result, the active edge configuration is preferred for efficient snapping. The size of the optimal active region scales with that of the natural boundary layer, $\sqrt{h/R}$ (Eq.~\eqref{thetaa_scaling}), with an empirical prefactor of about $2.5$ for our shells. 

We demonstrated the efficacy and generality of our findings using residually swelling and magneto-active shells. As we have shown, the curvature stimulus~\cite{Pezzulla2018} may be mapped to a broad range of non-mechanical loads, so we expect that these principles will apply widely to bistable actuators. While the homogenous curvature field assumed herein does not strictly apply to the magnetic stimulus, we found sufficient agreement with our experimental data. However, we expect that these and future results would be improved by more general theoretical framework that allows for a nonhomogenous curvature stimulus~\cite{Lee2021shelltheory}.

In our energy minimization scheme (Sect.~\ref{sect:optimize}), we assumed that the minimimum curvature stimulus was that for the fully active shell. This allowed us to identify the minimizing active area. However, we observed critical curvatures even below this value. We speculate that depending on the geometry, the torque at the active-passive interface can help to destabilize the shell--despite that it ostensibly acts in opposition to the edge torque. We leave investigation of this effect, which may open the door to further reduction of the critical curvature, for future work.

\begin{acknowledgements}
	We are grateful to Matteo Pezzulla for help with the COMSOL model, and to Kyle Hallock and Xin Jiang for helping to develop the experimental protocols. LSM, JHL, HSP and DPH gratefully acknowledge the financial support from NSF through CMMI-1824882. YY \& DPH gratefully acknowledge the financial support from Sartorius. ML, LSM \& DPH also acknowledge support from the Research in Science \& Engineering (RISE) program at Boston University. 
\end{acknowledgements}

\section{Appendix}

\subsection{Fabrication of non-homogenous residual swelling shells}\label{sect:shellfab}
\begin{figure}[h!]
	\centering	
	\includegraphics[width=0.94\linewidth]{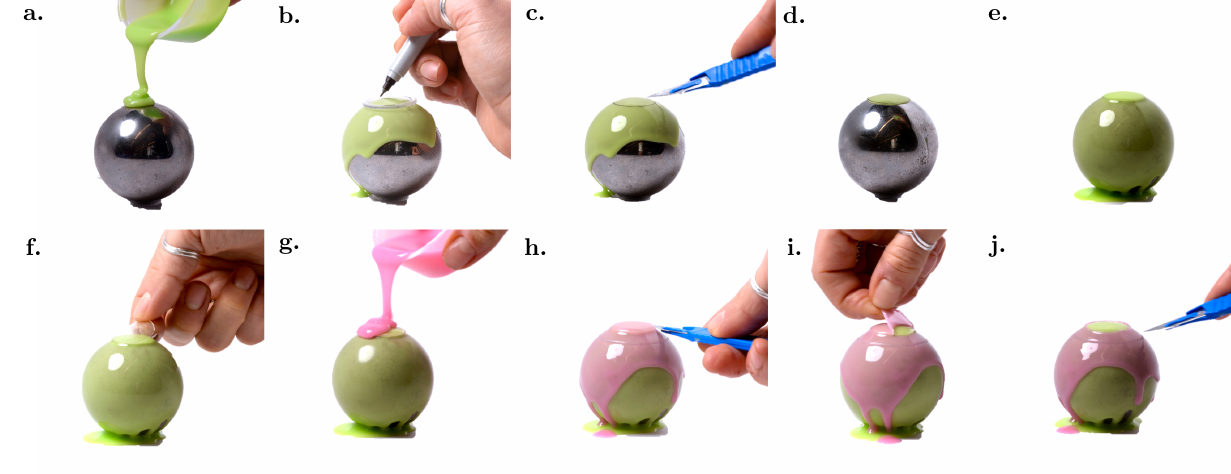}
	\caption{The fabrication process for a nonhomogeneous spherical cap with active edge: a. A layer of viscous green PVS is deposited over spherical metal ball-bearing. b. After curing, a ring of radius $R_p$ is centered at the top of the sphere and used as a stencil to mark $\theta_p$. c. A cut is made and d. excess material is removed, leaving a green layer of angular width $0\leq \theta \leq \theta_p$. e. A second green layer is added. f. Once the green layer is cured, a PDMS cap (cut from a sphere of the same size, in the same manner as a.-d.) of radius $R_p$ is centered at the top of the sphere. g. A pink layer of VPS is added. After curing, a shallow cut (h.) allows for the removal (i.) of the pink layer and the PDMS in the region $0\leq \theta \leq \theta_p$. j. A cut is made at $\theta$ (following a line which was stenciled by a ring as in b.) through all material to free the shell from the ball bearing. A small beam is cut vertically from just below the shell.}
	\label{fig:volvoxfab}
\end{figure}
The shell fabrication procedure for the active edge configuration is shown in Fig.~\ref{fig:volvoxfab}. The process is as follows: To begin, we coat a metal ball-bearing ($R_{\text{sphere}} \in [12,75]$ mm) with viscous polydimethylsiloxane (PDMS), ensuring a relatively uniform thickness~\cite{Lee2016}. Once the PDMS has cured, we use a laser-cut (Epilog Laser Helix, 75W) ring (inner radius $R_p \in [2,65]$ mm) as a stencil to guide a circular cut, resulting in a cap of opening angle $\theta_p = \theta-\theta_a = \sin^{-1}(R_p/R_{\text{sphere}})$. Next, we coat a ball bearing of the same size with green polyvinylsiloxane (PVS) (Fig.~\ref{fig:volvoxfab}a) and again make a cap with opening angle $\theta_p$. This time, the cap remains in place and the excess material is removed (Fig.~\ref{fig:volvoxfab}d). Since the active section necessarily has two layers of PVS, this additional passive layer ensures a relatively homogeneous thickness throughout the shell. With the green cap in place, a second layer of green PVS is added in the same manner (Fig.~\ref{fig:volvoxfab}e) and cut to the edge angle, $\theta \geq \theta_p$. It fuses completely to the first layer, so in the bulk (the eventual passive region, up to $\theta_p$) the material is thicker than at the edge at this stage of fabrication. After the second green PVS layer has cured, the PDMS cap is centered at the north pole (Fig.~\ref{fig:volvoxfab}f). It adheres to but does not fuse with the PVS. Next, we deposit a layer of pink PVS (Fig.~\ref{fig:volvoxfab}g). As soon as the pink layer is cured, a shallow cut is made around the edge of the PDMS cap (at $\theta_p$) (Fig.~\ref{fig:volvoxfab}h). Since the PDMS prevents crosslinking in the region it covers, we can peel the pink layer and the PDMS from this section (Fig.~\ref{fig:volvoxfab}i). Another laser cut ring ($R \in [6,75]$ mm, $R>R_p$) is used to guide a deeper cut through both layers, forming the bottom boundary (Fig.~\ref{fig:volvoxfab}j). This sets the total opening angle of the shell to be $\theta = \sin^{-1}(R/R_{\text{sphere}})$. We are left with a spherical cap composed of only green PVS (two layers thick) from opening angle $0$ to $\theta_p$, and a bilayer ring of angular width $\theta_a = \theta-\theta_p$ at the edge. In order to quantify the evolving curvature stimulus, we slice a small vertical beam from the material that remains on the ball-bearing just below the cut.

A shell with an active cap region is made in much the same way, except that the protective PDMS layer is a ring around the edge of the cap. This method prevents excision of a reliable bilayer beam from the region beneath the cut, so the curvature was not measured throughout the swelling process. Instead, a beam was cut from the cap region immediately following snapping. Because the curvature develops at a much slower rate than that of snap-through, the difference between the curvature immediately before and after snapping is negligible.

\subsection{Magnetic flux density as a curvature stimulus}\label{sect:flux-curvature}
\begin{figure}[h!]
	\centering
	\includegraphics[width = 0.55\linewidth]{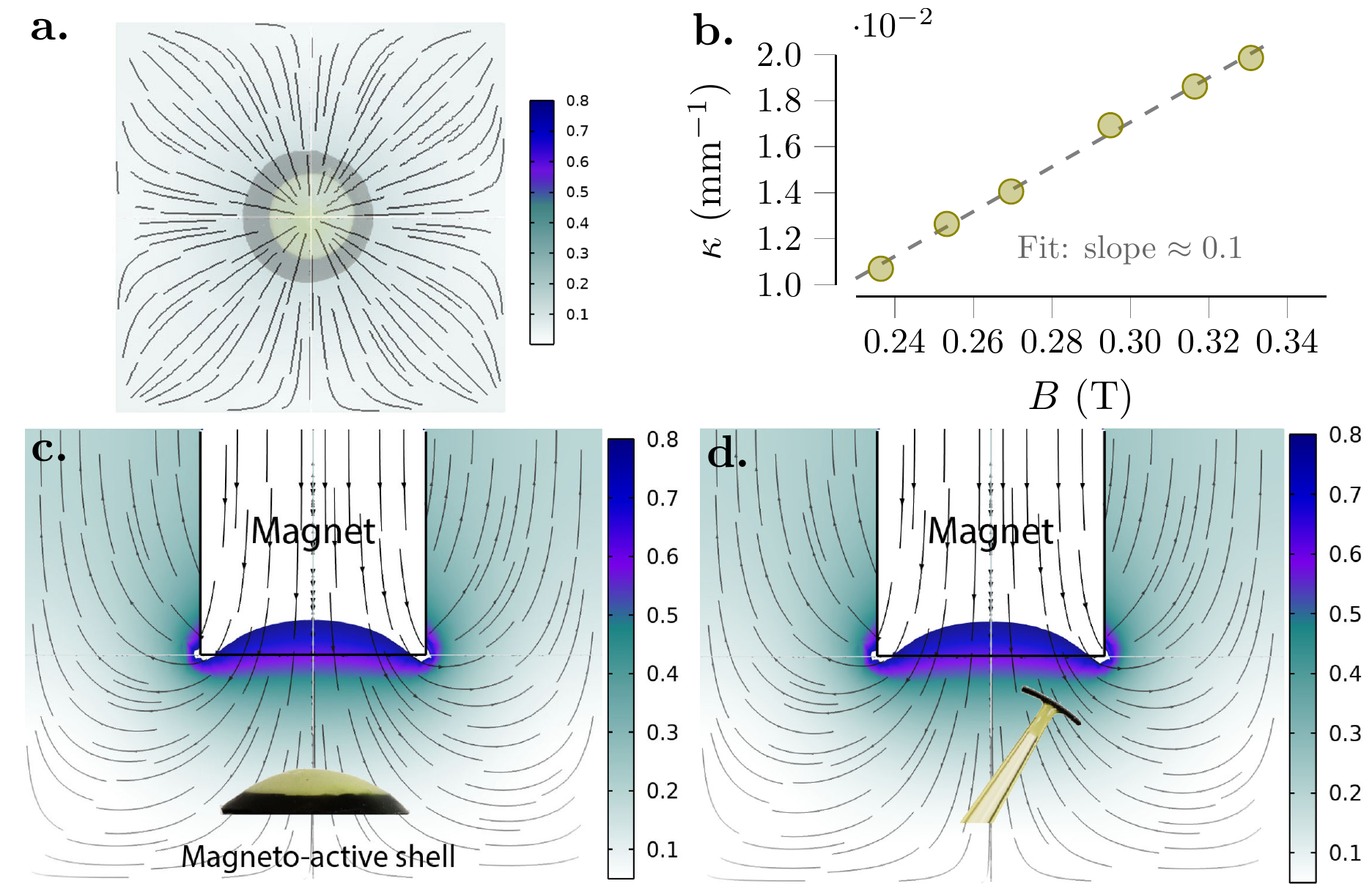}
	\caption{a. Finite element simulation showing the axially symmetric magnetic field generated from the NdFeB magnet. b. Calibration curve for magnetic flux density $\mathcal{B}$ \textit{vs.} the curvature stimulus $\kappa$, based on measurements taken from the beam in d. c. Magneto-active shell in the magnetic field. d. Magneto-active beam the magnetic field, which is oriented to match the edge of the shell and used to calibrate the $\kappa$--$\mathcal{B}$ relationship. In our range the relationship is approximately linear, and for our shells we used the empirical fit indicated by the dashed line. 
	}
	\label{fig:magneto}
\end{figure}

In order to visualize the magnetic field and obtain the magnetic flux density $\mathcal{B}$ at given spatial points, we simulate the magnetic field using the AC/DC module in COMSOL 5.6 (see Fig.~\ref{fig:magneto}). The finite element model is calibrated with experimental measurements and material properties provided by the manufacturer. The critical $\mathcal{B} = \mathcal{B}_c$ at the edge of the shell at the point of snapping was calculated by inputting the critical displacement (meaured from the Instron experiments) to the FE simulation.

To obtain the relationship between the applied magnetic flux density and corresponding natural curvature of the shell, we fabricated a beam of length $6.1$ mm, width $0.88$ mm, and thickness $h = 0.42$ mm. The beam is cut from the same spherical ball-bearing as the shells, so the initial curvatures are equivalent ($1/R=0.0787$ mm$^{-1}$). The arch is oriented at $0.639$ rad with respect to the vertical centerline to match the position of the corresponding ferromagnetic layer in the magnetic field (see Fig.~\ref{fig:magneto}c\&d.) The magnetic field is generated using the same cubic NdFeB magnet as is used for the shell. Digital image processing is used to measure the change of curvature as we vary the distance between the magnet and the bottom edge of the beam. The distance is then converted to magnetic flux density $\mathcal{B}$ using the COMSOL FE simulation. The response curvature as a function of the applied magnetic flux density over the range relevant to our experiments is shown in Fig.~\ref{fig:magneto}b. In our range, a first order polynomial fits the data. We note that the intercept of our empirical fit, which has a slope of $0.097$, is not zero, indicating that a linear fit would not be sufficient over a larger range.

 \newcommand{\noop}[1]{}

\end{document}